\documentclass[prl,twocolumn,showpacs]{revtex4-1}

\usepackage[utf8x]{inputenc}
\usepackage{amsmath,amssymb,amsfonts,subfigure,braket,color,dsfont}
\usepackage[english]{babel}
\usepackage{pdfpages}
\usepackage{subfig}

\usepackage{sistyle}

\usepackage{subfig}

\renewcommand{\H}{\mathcal{H}}

\DeclareMathOperator{\Var}{Var}

\begin{document}

\title{Hypothesis testing with open quantum systems}
\author{Klaus M{\o}lmer}
\email{moelmer@phys.au.dk}
\affiliation{Department of Physics and Astronomy, Aarhus University, Ny
  Munkegade 120, DK-8000 Aarhus C, Denmark.}

\date{\today}

\begin{abstract}
 Using a quantum circuit model we derive the maximal ability to distinguish which of several candidate Hamiltonians describe an open quantum system. This theory, in particular, provides the maximum information retrievable from continuous quantum measurement records, available when a quantum system is perturbatively coupled to a broadband quantized environment.
\end{abstract}


\maketitle

\nocite{Gill2000State,Gill2013From,Yuen1973Multipleparameter,Belavkin1976Generalized}

Two quantum states $\psi_0$ and $\psi_1$ can be distinguished unambiguously in a single experiment if they are orthogonal. If non-orthogonal states $\psi_0$, $\psi_1$ are provided with equal prior probabilities, the strategy distinguishing them with the smallest error probability performs a projective measurement on optimally chosen, orthogonal states $\tilde{\psi}_0$, $\tilde{\psi}_1$, (visualize a plane with the state vectors $\psi_0$, $\psi_1$ arranged symmetrically around the $45^\circ$ direction between orthogonal vectors $\tilde{\psi}_0$ and $\tilde{\psi}_1$). The state vector overlap $\alpha=\langle \psi_1|\psi_0\rangle$ specifies $|\langle\psi_\theta|\tilde{\psi}_\theta\rangle|^2=\frac{1}{2}(1+\sqrt{1-|\alpha|^2})$, $\theta=0,1$, and the optimal guess that the prepared state was $\psi_\theta$ if one measures $\tilde{\psi}_\theta$, has an error probability of
\begin{equation}
P_e=\frac{1}{2}(1- \sqrt{1-|\alpha|^2}). \label{Perror}
\end{equation}

Discrimination of quantum states is related to hypothesis testing \cite{Tsang} and parameter estimation \cite{Braunstein}. To determine if the evolution of a quantum system is governed by one or another Hamiltonian, one must perform measurements on the system and use their outcome to infer which is the most likely assumption. We are interested in the situation of an open quantum system, S, whose  interaction with a broad band environment, E, permits the Born-Markov approximation. \textit{I.e.}, continuous monitoring of the environment as depicted in Fig. 1a) does not alter the relaxation properties of the system (no quantum Zeno-effect). In \cite{Tsang} it was shown how to discriminate different hypotheses optimally from a given measurement record by solution of a conditional master equation, and, e.g., \cite{Wiseman,Mabuchi,particle,GammelmarkBayes,Negretti,TsangPRA,TsangNJP,waveform} have investigated strategies to obtain precise estimates of physical parameters and timedependent excitation waveforms from detection signals.

Photon counting and field quadrature measurements represent different ways to probe an optical field with correspondingly different stochastic master equations \cite{mcwf,carmichael,WisemanBook}. Rather than  addressing particular measurement schemes, we present a method to evaluate the optimal discrimination allowed by \textit{any} monitoring of the environment of the system (the emitted radiation) and the final state of the system itself.
Such an analysis is possible because, under validity of the Born-Markov approximation, a sequence of measurements on the environment  may be deferred to a final measurement of appropriate environment degrees of freedom: The temporal sequence of detector clicks associated with counting of photons emitted by an atom during time $[0,t]$, is for example equivalent to the counting at time $t$ of photons in volume elements at corresponding distances from the atom.

\begin{figure}
\centering
\includegraphics[page=1,trim=0 220 0 0,width=6truecm,]{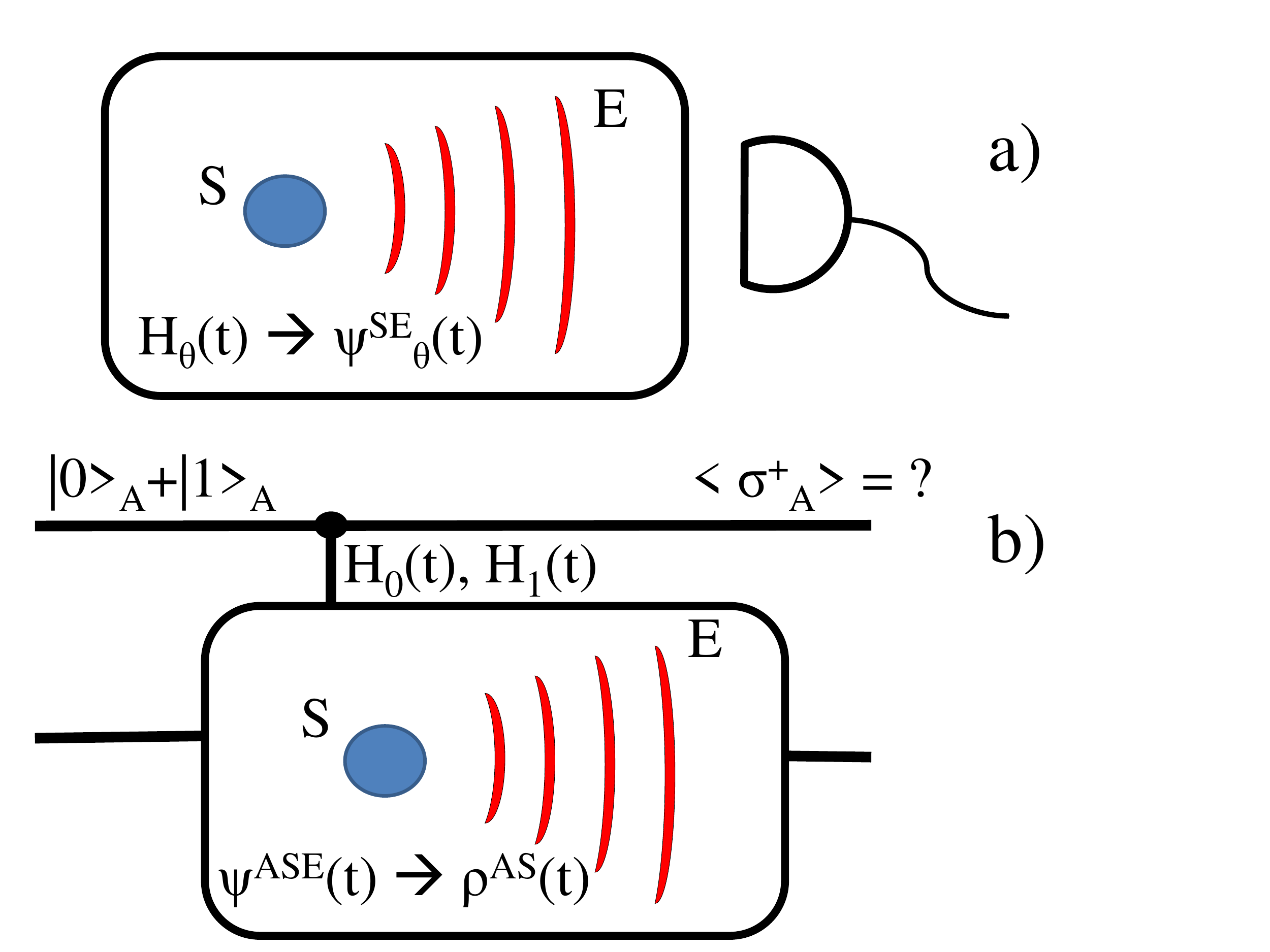}
\vspace{2.5cm}
 \caption{Panel a) illustrates the evolution of a quantum system, S, and a broadband environment, E, governed by a Hamiltonian $H_\theta(t)$. In b), the state of an ancilla qubit controls application of candidate Hamiltonians $H_0(t),\ H_1(t)$. The reduced system and ancilla density matrix $\rho^{AS}$ follows by elimination of the environment.}
\label{fig:fig1}
\end{figure}

The information retrieved by measurement records is bounded by our ability to discriminate unprobed states $\psi_0^{SE}(t)$ and  $\psi_1^{SE}(t)$ of the system and environment which, according to (\ref{Perror}), is given by $\alpha_{SE}=\langle \psi_1^{SE}(t)|\psi_0^{SE}(t)\rangle$. Evaluation of the joint quantum state of the system and environment is prohibitively complicated as it requires inclusion of a vast number of photon number states distributed in an entangled manner over a continuum of field modes. The state vector overlap, however, can be obtained without recourse to calculation of the states.

The principle behind our calculation of $\alpha_{SE}$ is illustrated in Fig.1. In part (a) of the figure we sketch the quantum system and its environment subject to the Hamiltonian $H_\theta$ and, possibly, to continuous probing of the radiation emitted into the environment. In part (b) of the figure, we introduce the ancilla A and the ancilla qubit-controlled Hamiltonian $H_{ASE}=(|0\rangle\langle 0|)_A\otimes H_0(t)+(|1\rangle\langle 1|)_A\otimes H_1(t)$ which represents two different candidate Hamiltonians $H_0(t)$ or $H_1(t)$. Such inclusion of ancilla qubit degrees of freedom has been proposed for a variety of tasks, including quantum computing on mixed state quantum systems with one pure qubit \cite{DQC1}, estimation of entanglement \cite{Replica} and thermodynamical \cite{Phase} properties. Similar, higher dimensional ancillary degrees of freedom are used in particle filter theory with stochastic master equations \cite{particle,Negretti}. We emphasize that the ancillary qubit is merely introduced as a theoretical construction to represent alternative hypotheses in a convenient manner.

The ancilla, system and environment are initially prepared in a pure state $\frac{1}{\sqrt{2}}(|0\rangle_A + |1\rangle_A)\otimes |\psi ^{SE}(t=0)\rangle$, which evolves into
\begin{eqnarray} \label{psiASE}
|\psi(t)\rangle  =  \frac{1}{\sqrt{2}}\bigl(|0\rangle_A\otimes |\psi_0^{SE}(t)\rangle + |1\rangle_A)\otimes |\psi_1^{SE}(t)\rangle\bigr).
\end{eqnarray}
Note that the desired wave function overlap $\alpha_{SE}=\langle \psi_1^{SE}(t)|\psi_0^{SE}(t)\rangle$ can be formally evaluated as twice the expectation value of the raising operator, $\sigma^+_A  = (|1\rangle \langle 0|)_A$, of the ancilla qubit:
\begin{equation} \label{overlapstep}
\langle\psi_1^{SE}(t)|\psi_0^{SE}(t) \rangle = 2\langle \sigma^+_A \rangle.
\end{equation}

At this point we use our assumption that the Born-Markov appoximation applies for the system-environment interaction, such that the environment degrees of freedom can be eliminated. We write the density matrix of the ancilla and the system as the following $2\times 2$ matrix of matrices
\begin{equation} \label{bigm}
\rho^{AS}(t) = \frac{1}{2} \left(
              \begin{array}{cc}
                \rho_{00}(t) & \rho_{01}(t) \\
                \rho_{10}(t) & \rho_{11}(t) \\
              \end{array}
            \right),
            \end{equation}
where $\rho_{\mu\nu}$, acting on the system state space, are initially identical, $\rho_{\mu\nu}(0)=\rho^S(0) = |\psi^S(0)\rangle\langle\psi^S(0)|$.

The rows and columns in (\ref{bigm}) correspond to the different ancilla states which cause the evolution of the system and the environment under different Hamiltonians. If the hypotheses concern only the unitary part of the system evolution, we apply the ancilla and system Hamiltonian,
\begin{equation}
\H^{AS} =  \left(
              \begin{array}{cc}
                H^S_0(t) & 0 \\
                0 & H^S_1(t) \\
              \end{array}
            \right),
            \end{equation}
while, to represent different environment couplings (e.g., hypotheses concerning different strengths or different system relaxation operators, the ancilla and the system are subject to Lindblad relaxation terms, $\dot{\rho}^{AS} = \sum_m {\cal D}[\hat{c}^{AS}_m ]\rho^{AS}$, where ${\cal D}[\hat{c}]\rho\equiv \hat{c}^\dagger \rho \hat{c}- \frac{1}{2}(\hat{c}^\dagger \hat{c}\rho + \rho \hat{c}^\dagger \hat{c})$, and where $\hat{c}^{AS}_m$ is of the form:
  \begin{equation} \label{damp}
\hat{c}^{AS}_m =  \left(
              \begin{array}{cc}
               \hat{c}^{0}_m  & 0 \\
                0 & \hat{c}^{1}_m \\
              \end{array}
            \right).
\end{equation}

It follows from (\ref{overlapstep},\ref{bigm}) that the desired overlap is given by the trace $\alpha_{SE}=$Tr$_S(\rho_{01})$. The matrix $\rho_{01}$ is subject to the combined action of the candidate Hamiltonians and relaxation terms, and solves the equation,
\begin{eqnarray} \label{twosided}
\dot{\rho}_{01} = \frac{1}{i\hbar}(H^S_0 \rho_{01} - \rho_{01} H^S_1) + \nonumber \\
\sum_m \hat{c}^{0}_m \rho_{01} (\hat{c}^{1}_m)^\dagger - \frac{1}{2}((\hat{c}^{0}_m)^\dagger \hat{c}^{0}_m \rho_{01} + \rho_{01} (\hat{c}^{1}_m)^\dagger \hat{c}^{1}_m).
\end{eqnarray}
This equation is structured like the Lindblad master equation, but all operators multiplying $\rho_{01}$ from the left (right) pertain to hypothesis 0 (1). Unlike the usual master equation which conserves the trace of the density matrix, the left and right multiplication with different Hamiltonian and Lindblad operators break this invariance and cause non-trivial time evolution of $\alpha_{SE}$. An alternative derivation of  Eq.(\ref{twosided}) using quantum measurement theory was applied in \cite{Madalin,GammelmarkCRB,Madalin2}. The present derivation is more straightforward and allows generalization to a wider range of problems.

For illustration, consider a two-level atom with a ground state $\ket g$ and excited state $\ket e$ driven on resonance with a Rabi-frequency $\Omega_0=0$ or $\Omega_1=4\kappa$ while the excited state decays by fluorescence emission with a rate $\kappa$. The atom is initialized in its ground state at $t=0$. It is straightforward to solve Eq.(\ref{twosided}) with different $H^S_\theta = \frac{\hbar \Omega_\theta}{2}(|e\rangle\langle g|+|g\rangle\langle e|)$ and identical Lindblad damping operators $\hat{c}^\theta=\sqrt{\kappa}|g\rangle\langle e|$, and in Fig. 2, we show with the fat solid curve the error probability according to (\ref{Perror}) with $|\alpha_{SE}|^2 =|$Tr($\rho_{01}(t))|^2$. This curve yields the ultimate limit to our ability to discriminate among Rabi frequencies $\Omega_0$ and $\Omega_1$.

The detection of just a single photon is incompatible with $\Omega=0$, while if no photon is detected our best guess among the two choices is that $\Omega=0$. If $\Omega=\Omega_1$, the probability of detecting no photons until time $t$ can be obtained by propagating the so-called no-jump wave function of the system, $|\psi^S_{\textrm{NJ}}(t)\rangle = a(t)|g\rangle + b(t)|e\rangle$, according to a non-hermitian system Hamiltonian , $H_{\textrm{NH}}=\frac{\hbar\Omega_1}{2}(|e\rangle \langle g|+|g\rangle \langle e|) - \frac{i\hbar\kappa}{2}|e\rangle\langle e|$, and the probability of observing no photon detection event is given by $P_{\textrm{NJ}}=||\psi^S_{\textrm{NJ}}(t)||^2$ \cite{mcwf,carmichael}. The probability that $\Omega=\Omega_1$ is wrongly associated with $\Omega=0$ is thus $\frac{1}{2}(|a(t)|^2+|b(t)|^2)$, shown as the thin solid curve in Fig. 2 . From an initial value of $\frac{1}{2}$ the error probability decreases, as it becomes less and less likely that no photon has been detected from the laser driven atom.

Had we instead considered the case, where only a measurement on the atom is allowed at the end of the interaction time $t$, such a measurement should distinguish between the two density matrices, $\rho_0$ and $\rho_1$, evolved by the Lindblad master equations with the different Hamiltonians. The minimal probability of making an assignment error is here provided by Helstrom \cite{Helstrom}, $P_e^\rho = \frac{1}{2} + \sum_{\gamma_j \leq 0} \gamma_j$, where the sum is over the negative eigenvalues of the operator $\frac{1}{2}(\rho_1-\rho_0)$. In Fig. 2, $P_e^\rho$ is shown as the dash-dotted green curve for the case of $\Omega_0=0$ and $\Omega_1=4\kappa$. Since the system evolves into steady states with only partially distinguishable density matrices, the error probability by atomic detection does not approach zero in the long  time limit.

The quantity $|a(t)|^2$ derived above for the un-normalized no-jump wave function is the probability that, despite the non-vanishing $\Omega=\Omega_1$, no photon has been detected and the atom is in its ground state at time $t$. Using photon counting \textit{and} detection of the final atomic excitation thus yields the erroneous assignment (of a vanishing Rabi-frequency when $\Omega=\Omega_1$) with a probability $|a|^2/2$, shown as the dashed red curve in Fig. 2. We observe that at particular finite probing times, we can distinguish the hypotheses with certainty. These are the times where the no-jump wavefunction has no ground state population, and a non-vanishing Rabi frequency results in a photonic or atomic excitation with certainty.

The probability $|a(t)|^2$ for observing  no photon and no atomic excitation equals the population of the quantum state component with no atomic or field excitations. This is precisely the state $|\psi^{SE}_0(t)\rangle$ obtained for $\Omega_0=0$ and, hence the overlap between the candidate system and environment states is given by, $|\alpha_{SE}|^2=|a(t)|^2$. We note that although the photon counting analysis permits calculation of the minimal error probability (\ref{Perror}), the optimal measurement achieving this error is more complicated as it involves projection on entangled superposition states $\{|\tilde{\psi}_\theta^{SE}\rangle\}$ of the atom and the quantized radiation field.

\begin{figure}
\centering
\includegraphics[page=1,trim=60 230 60 230,width=6truecm,]{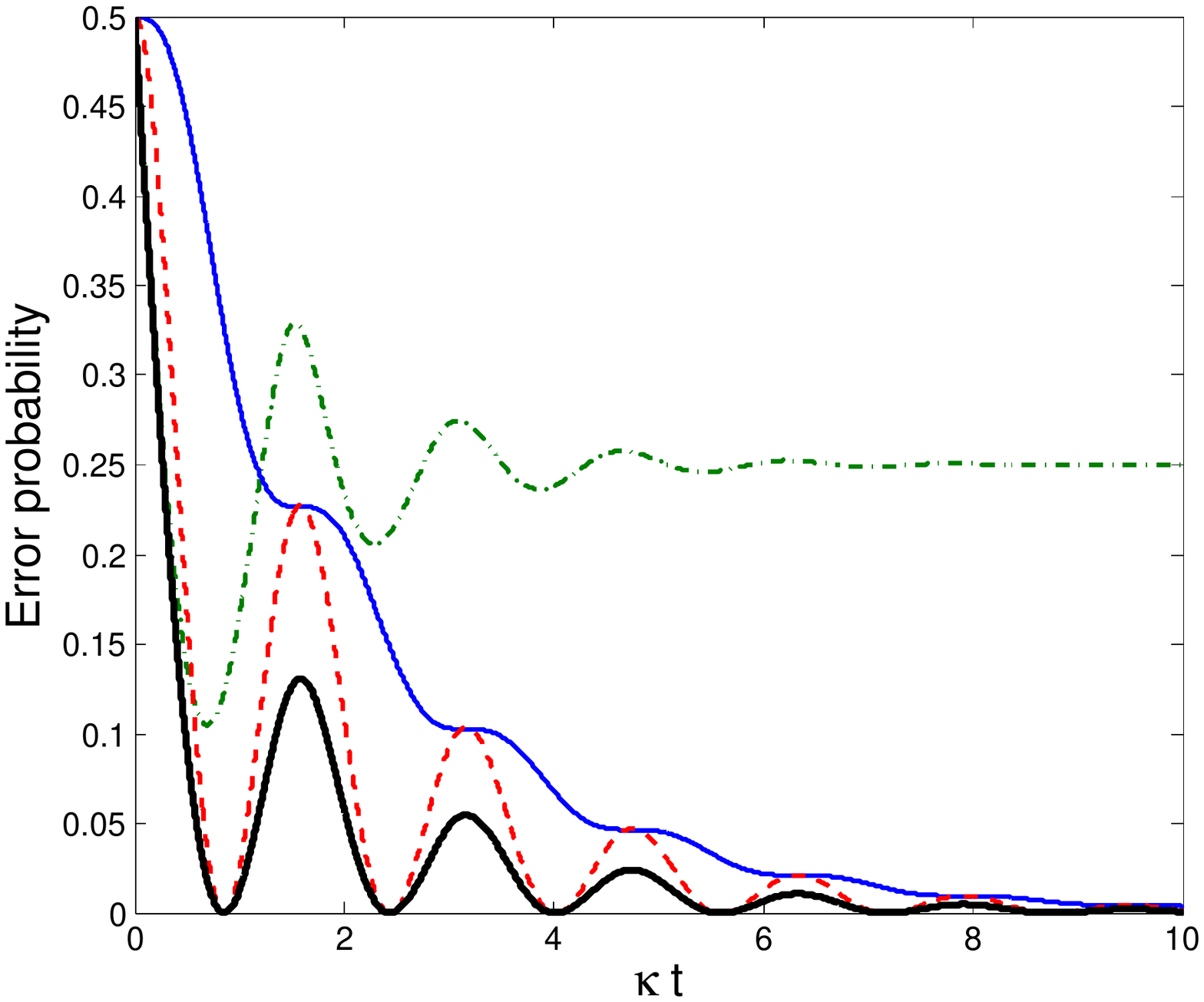}
 \caption{(Color online) Time dependent probability for erroneously assigning whether an atom with decay rate $\kappa$ is excited resonantly with a Rabi frequency of $\Omega_0=0$ or $\Omega_1= 4\kappa$. The error probability is shown for a purely atomic measurement (green, dash-dotted curve), photon counting (blue, thin solid curve), photon counting and a measurement of the final atomic excitation (red, dashed curve). The black, fat solid curve shows the minimal error achievable by any measurement on the system and the radiation field.}
\label{fig:fig2}
\end{figure}

Eq.(\ref{twosided}) is readily solved for any sets of hypotheses about Rabi frequencies, and we find that, for a vanishing detuning, our ability to distinguish two real Rabi frequencies depends only on their difference $\Omega_1-\Omega_0$. This is because addition of an extra driving  Hamiltonian which commutes with all the other Hamiltonian terms causes a common unitary rotation and hence no change of the overlap of the system and environment states, $\psi_\theta^{SE}$. Despite the striking fact that candidate values of same strength but opposite sign, $\Omega_0=-2\kappa,\ \Omega_1=2\kappa$, yield equivalent photon count signals and final atomic excited state populations, according to Eq.(\ref{twosided}), they are equally well distinguished as $\Omega_0=0,\ \Omega_1=4\kappa$. Extra Rabi frequency terms do not, however, commute with detuning terms in the Hamiltonian, and we obtain different results when the system is excited off resonance.

One may readily imagine strategies to improve experiments to obtain faster or stronger discrimination. Our theory constitutes an excellent starting point for such an optimization effort, varying, e.g., the initial state and available control Hamiltonians added to both $H_0$ and $H_1$ in (\ref{twosided}) with the aim to minimize $|$Tr$(\rho_{01}(t)|^2$. As an example of such optimization, we have considered the ability to vary the Rabi frequency in experiments aiming to distinguish whether a two level system is driven on resonance or with a given finite detuning $\delta$ (caused, e.g., by dispersive coupling to an external influence). After a fast transient, the error probability for this assignment decays exponentially with time, and since Eq.(\ref{twosided}) is a linear set of equations, we can find the characteristic time scale of this decay from the eigenvalue of the corresponding $4\times 4$ matrix with the smallest (negative) real part. Since very weak excitation yields no fluoresence signal while very strong excitation causes power broadening of the transition, we expect that there exists an optimum Rabi-frequency. The curves in Fig. 3 for different detunings that we want to distinguish from zero show the smallest negative real part of the eigenvalues $\lambda$ under variation of the Rabi frequency between $0$ and $1.5\kappa$. They all vanish for $\Omega \rightarrow 0$, while their largest value, and hence the fastest convergence of $\alpha_{SE}$, occurs for intermediate values of $\Omega \sim 0.75\kappa$, except for $\delta=\kappa/2$ which is ideally distinguished from zero by a weaker driving field, $\Omega=0.62\kappa$. Use of a time dependent $\Omega(t)$  constitutes an attractive possibility to further explore and optimize the discrimination of different detunings.

\begin{figure}
\centering
\includegraphics[page=1,trim=60 230 60 230,width=6truecm,]{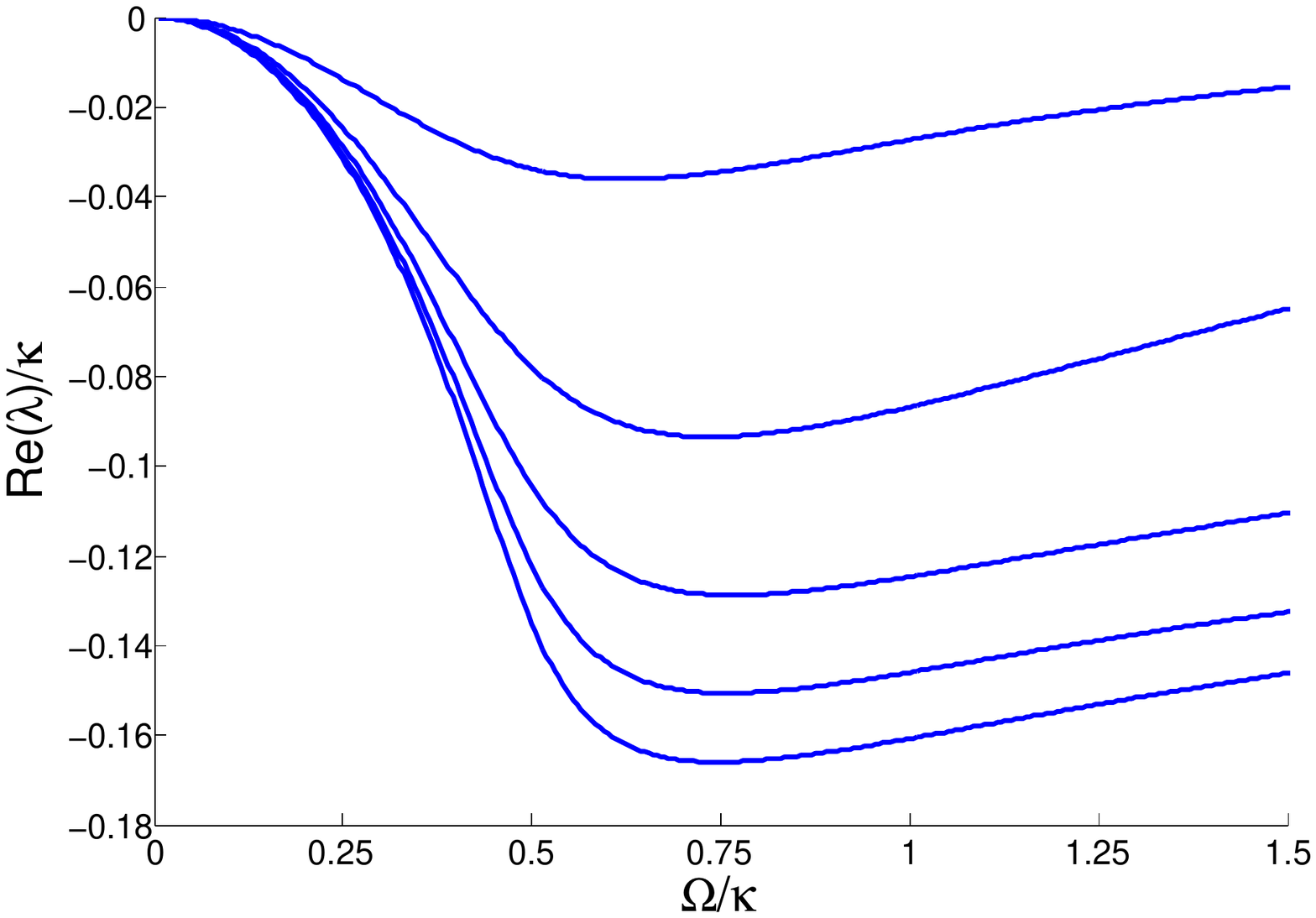}
 \caption{The numerically smallest real part of the eigenvalues of Eq.(\ref{twosided}) governing an exponentially converging distinction of different discrete values of the detuning $\delta$ from zero. The values of the convergence rate are shown as function of the Rabi frequency $\Omega$ for $\delta=0.5\kappa, \kappa, 1.5\kappa, 2\kappa, 2.5\kappa$ (from above).}
\label{fig:fig2}
\end{figure}

Our theory also encompasses probe master equations \cite{WisemanBook}, describing e.g., dispersive phase-shift or polarization-rotation of an optical field due to its interaction with a quantum system. In the probe master equation, normally a stochastic back action term appears, depending on the kind of measurement performed on the probe field and its efficiency. Our two-sided master equation omits the random back action terms, and $|$Tr$(\rho_{01}(t)|^2$ yields the optimum information gain by any probe measurement. Variation of the probe field strength, detuning and polarization corresponds to variation of the terms $\hat{c}_m$, and entails possibilities to optimize the information gain.

Our use of examples with a simple two-level atom does not restrict application of our theory to cases where a Lindblad-like master equation  Eq.(\ref{twosided}) can be solved for the matrix element of $\rho_{01}$. After introduction of an ancilla to encode different hypotheses, the quantity $\langle \sigma_A^+ \rangle$ can be treated as a conventional physical observable, and any theory that permits its calculation may be applied. We can, e.g., simulate the time-evolution of (\ref{bigm}) and determine $\langle \sigma_A^+ \rangle$ by sampling with Monte Carlo wavefunctions \cite{mcwf,carmichael}, and while these functions may correspond to a particular (photon counting) experiment, the evaluation of the average $\langle \sigma_A^+ \rangle$ yields the limit of discrimination by any detection scheme.

While the practical availability of the information emitted into the environment is compatible with Markovian decay, non-Markovian master equations can be sometimes derived for certain system-environment models. If, e.g., the steps leading to a time convolutionless master equation \cite{Haikka}) can be carried out for the ancilla-augmented system, its solution  will provide an upper limit to the discriminating power based on the probing of the system and environment at a given final time. Recall, however, that the non-Markovian dynamics may not be compatible with continuous probing and measurement back action until that time.

We are also not restricted to treatments in any definite representation of the quantum system, and we may apply evolution in the Heisenberg picture, input-output theory, phase space distribution functions, and, when applicable, Gaussian covariance matrices \cite{Gauss,TsangPRA,Anne}. While open many-body problems may not be generally tractable, and one may have recourse to numerical studies on finite systems, perturbative or variational methods may apply in special cases to evaluate expectation values with sufficient precision. Some many-body systems may thus be adequately described by  Hartree-Fock or multi-orbital mean-field theory \cite{Cederbaum}, and matrix product states \cite{Verstraete2008a}, to mention a few approximate treatments.

To offer an example, a large ensemble of two-level systems, probed off resonance, can be described by collective spin variables well approximated by canonical conjugate variables $\hat{x},\hat{p}$ and a Gaussian coherent initial state \cite{Gauss}. Rotation of the spins due to different candidate magnetic fields is represented by Hamiltonian terms $\hbar g_i \hat{p}$ that cause different coherent displacements $D(g_i t)$. Let $k$ denote the strength of the probe term $-k[\hat{x},[\hat{x},\rho]]$ in the master equation. The two-sided interaction picture ansatz, $\rho_{01}(t) = D(g_0 t)\sigma(t)D(-g_1 t)$, yields the equation $\dot{\sigma}= -k\bigl((\hat{x}-g_0 t)^2\sigma + \sigma(\hat{x}-g_1 t)^2 - 2(\hat{x}-g_0 t)\sigma(\hat{x}-g_1 t)\bigr)$, and in the continuous $x$-representation, $\sigma(x,x',t)$ is readily found by direct integration over time of an $(x,x')$-dependent exponential factor. The final result for Tr$(\rho_{01}(t)) = e^{-(g_0-g_1)^2 t^2/4}\cdot e^{-(g_0-g_1)^2 k t^3/3}$ is interesting: The first factor yields the overlap of the displaced coherent states available for a final measurement whether $k$ vanishes or not. The second factor shows how probing for time $t$ entangles the spins by gradually squeezing their collective spin variable. The overlap therefore converges faster than the exponential dependence discussed above for a single system, and the sensitivity increases. In \cite{Gauss}, we, indeed, found that homodyne detection allows $B$-field estimation with an error scaling as $1/t^3$, in agreement with the scaling of the discrimination error following from our expression for Tr$(\rho_{01}(t))$.

In conclusion, we have used a circuit model with a qubit ancilla to address hypothesis testing. We have identified a simple reduced system  operator and an associated effective master equation that yield the scalar product between pure quantum states of the system and the environment. This scalar product sets the limit to how well the states, and hence the evolution hypotheses, may be distinguished by any measurement scheme. Our theory may guide efforts in the search for efficient practical schemes, and since optimal distinguishability is achieved by projection on entangled states of the system and the environment, it may be interesting to analyze adaptive schemes that choose among measurements according to earlier detection outcomes \cite{Wisemanadapt,Mabuchiadapt,TsangPRA}.

While we have given examples of testing between distinctly different hypotheses, our theory also allows estimation of the \textit{precision} by which an unknown continuous parameter can be determined. According to \cite{Braunstein}, the estimation error on a continuous parameter $\theta$ scales asymptotically according to the Cram\'{e}r-Rao bound $\Var(\hat\theta) \geq \frac{1}{I(\theta)}$, where $I(\theta)$ is the Fisher information,
$I(\theta) = 4\Re(\braket{\partial_\theta \psi(\theta)|\partial_{\theta'} \psi(\theta')} - \braket{\partial_\theta\psi(\theta) |\psi}\braket{\psi|\partial_{\theta'}\psi(\theta')})_{\theta=\theta'}$.
Our theory provides the scalar products between states and - by a finite difference approximation - the derivatives needed to evaluate $I(\theta)$. An alternative, perturbative approach to obtain the derivatives is illustrated in the Supplementary Material of Ref.\cite{GammelmarkCRB} and, for a different problem, in \cite{Madalin2}.

The effective evaluation of our theory makes it a good starting point for optimization and for studies of the role of finite detection efficiency and unobserved dissipation channels \cite{TsangNJP,GammelmarkBayes}. It may also provide crucial insights into the consequence of, e.g., measurement feedback, phase transitions and large deviation behaviour \cite{TsangPRA,gammelmarkphase,tsangphase,newphase,largedev} for hypothesis testing and parameter estimation.

Discussions with Alexander Holm Kiilerich and support from the Villum Foundation are gratefully acknowledged.

\end{document}